\documentclass[prd,aps,preprintnumbers,amssymb]{revtex4}
\usepackage{axodraw}
\usepackage{color}
\usepackage{epsf}

\begin{document}
\title{%
\hfill{\normalsize\vbox{%
\hbox{}
 }}\\
{Supersymmetric QCD with $N_f<N$ revisited }}

\author{Renata Jora
$^{\it \bf a}$~\footnote[2]{Email:
 rjora@theory.nipne.ro}}

\affiliation{$^{\bf \it a}$ National Institute of Physics and Nuclear Engineering PO Box MG-6, Bucharest-Magurele, Romania}

\date{\today}

\begin{abstract}
We introduce in supersymmetric QCD with $N_f<N$ additional meson singlet degrees of freedom made of $2N_f-2$ supermultiplets. We construct an alternative superpotential with $N_f^2$ constraints and show its consistency. Based on this superpotential we prove that for $\frac{N}{2}<N_f<N$ the supersymmetry is unbroken whereas for $N_f\leq\frac{N}{2}$ is dynamically broken.
 
\end{abstract}

\maketitle

\section{Introduction}
The non-perturbative behavior of the most common supersymmetric gauge theories was analyzed and elucidated  by Seiberg and his collaborators in  a series of groundbreaking works \cite{ADS}, \cite{Seiberg1}, \cite{Seiberg2}. Among the discussed cases was also $SU(N)$ supersymmetric gauge theory with $N_f$ matter supermultiplets in the fundamental and antifundamental representations.  This model is of the utmost importance because of the similarities with the more mundane QCD.

By using the tools of holomorphicity and dualities between different regimes   the behavior of the theory as a function of the number of flavors $N_f$ was elucidated by pointing out when various phases occur and when chiral symmetry breaking takes place. It is known that in general for suppersymmetric QCD  supersymmetry is not dynamically broken except for the case $N_f< N$ which is problematic. If a supersymmetric gauge theory displays or not dynamical supersymmetry breaking is crucial for the construction of supersymmetric extensions of the standard model of elementary particles.   In most cases  an additional supersymmetry breaking  sector must be introduced through soft terms.

The case $N_f<N$ is critical. The theory has a global symmetry $SU(N_f)_L\times SU(N_f)_R\times U(1)_B\times U(1)_R$ with a squark assignment under $U(1)_B \times U(1)_R$ of ($1, \frac{N_f-N}{N_f}$) (and of anti squarks ($-1,\frac{N_f-N}{N_f}$)). The quarks assignment is ($1,-\frac{N_f}{N}$) (with the antiquarks assignment ($-1,-\frac{N}{N_f}$)).  There are in general $D$ flat directions in the moduli space. The vacuum expectations values for the squarks can be brought into the diagonal form:
\begin{eqnarray}
\langle \tilde{\Phi}^*\rangle =\langle \Phi\rangle =
\left(
\begin{array}{ccccc}
v_1&0&0&...&0\\
0&v_2&0&...&0\\
0&0&0&...&v_f\\
...&...&...&...&...\\
0&0&0&...&0\\
0&0&0&...&0
\end{array}
\right).
\label{assign45534}
\end{eqnarray}
where the matrix has the dimension $N\times N_f$.

It can be shown \cite{ADS} \cite{Seiberg1}, \cite{Seiberg2} that in the low energy regime the theory is described by the $ADS$ superpotential:
\begin{eqnarray}
W_{ADS}=(N-N_f)\Bigg(\frac{\Lambda^{3N-N_f}}{\det M}\Bigg)^{\frac{1}{N-N_f}},
\label{ADsuperpot66645}
\end{eqnarray}
where $\Lambda$ is the holomorphic scale of the theory and $M$ is a $N_f\times N_f$ matrix field that describes the uneaten chiral supermultiplets $M^j_i=\tilde{\Phi}^{jn}\Phi_{ni}$.
Is then evident from Eq. (\ref{ADsuperpot66645}) that the scalar potential is zero for $M=\infty$. Then the theory may be assumed with dynamical supersymmetry breaking because the zero potential is obtained or without dynamical supersymmetry breaking because the zero potential is obtained for $M\rightarrow \infty$. One says that there is no ground state of the theory. It is known that if supersymmetry is dynamically broken for some values of the coupling constant it is in general broken. Moreover in this case the Witten index  $(-1)^F=n_{b0}-n_{F0}$ \cite{Witten} which is given by the difference between the bosonic zero modes and fermionic modes should be zero.

At this point we need to consider what through decades of study one can learn from regular QCD. Below some scale the theory confines and bound states of quarks, the mesons and the baryons form. Moreover it is perfectly possible that the mesons and baryons may contain more than the common number of quarks which are two for mesons and three for baryons. In particular bound states of tetraquark mesons may exist \cite{Jaffe}, \cite{Jora}.  This is in general a good assumption that helps explain the unusual inverted mass spectrum of the scalar mesons.

Then it is natural to wonder: what happens if in the low energy description of supersymmetric QCD one introduces additional degrees of freedom, mesons states composed of $2N_f-2$ squarks (and the corresponding quark structure). Could this help elucidate some of the aspects still unclear regarding the behavior of the theory for $N_f<N$?
In this work we will show that although this idea might seem artificial and redundant it can properly and  consistently be implemented in the supersymmetric QCD with $N_f<N$. Moreover in this context one can show without any doubt that the theory has the supersymmetry unbroken for $N_f>\frac{N}{2}$ whereas for $N_f\leq\frac{N}{2}$ the supersymmetry is broken. Section II contains the description of the states and the alternative superpotential that can be constructed out of these. In section III we show in which cases supersymmetry is dynamically broken and in which it is not. The conclusions are drawn in section IV.

\section{The superpotential}

We introduce additional meson fields made of $2N_f-2$ squarks:
\begin{eqnarray}
M^{\prime\prime  i_1}_{j_1}=\epsilon^{i_1i_2...i_{N_f}}\epsilon_{j_1j_2...j_{N_f}}M^{\dagger j_2}_{i_2}...M^{\dagger j_{N_f}}_{i_{N_f}}.
\label{res534428}
\end{eqnarray}

One can construct many holomorphic invariant out of $M$ and $M'$ but it is sufficient to consider only two of them:
\begin{eqnarray}
&&I_2={\rm Tr}MM^{\prime\prime \dagger}
\nonumber\\
&&I_3=\det[M^{\prime \prime \dagger}].
\label{finalres554663}
\end{eqnarray}
Note that by construction these invariants are still holomorphic. For the simplicity of the notation we redefine $M'=M^{\prime\prime\dagger}$.

We need to match the correct degrees of freedom. It is known that for $N_f<N$ there are generically $N_f^2$ matter degrees of freedom left is the gauge group is broken down to $SU(N-N_f)$. Then if we introduce two matrices $M$ and $M'$ there are $2N_f^2$ degrees of freedom and one needs $N_f^2$ constraints. We will consider these as obtained from the equation of motion for the fields $M$. Then one needs to introduce in the superpotentail the combination $tI_1-I_2$ where $t=(N_f-1)!$. One can check,
\begin{eqnarray}
\frac{\partial (tI_1-I_2)}{\partial M^{i_1}_{j_1}}=\epsilon^{i_1i_2...i_{N_f}}\epsilon_{j_1j_2...j_{N_f}}M^{ j_2}_{i_2}...M^{ j_{N_f}}_{i_{N_f}}-M^{\prime i_1}_{j_1}=0.
\label{constr66453}
\end{eqnarray}
This leads to the $N_f^2$ wanted constraints.

We will present here the final version of the superpotential we propose obtained through matching the correct behavior under anomalies and the correct quantum numbers by analogy with the construction of the $ADS$ superpotential:
\begin{eqnarray}
W=(N-N_f)(I_1t-I_2)^{-\frac{N_f}{N-N_f}}I_3^{\frac{1}{N-N_f}}(\Lambda^b)^{\frac{1}{N-N_f}}(-1)^{\frac{N_f}{N-N_f}}N_f^{\frac{N_f}{N-N_f}}=H(I_1t-I_2)^{-\frac{N_f}{N-N_f}}I_3^{\frac{1}{N-N_f}}.
\label{superpot64553}
\end{eqnarray}
Here $I_1=\det M$. 

It is obvious that the potential in Eq. (\ref{superpot64553}) is similar in many ways with the $ADS$ superpotential and that satisfies all the requirements. 

Here we will check only the consistency of the superpotential in the case one flavor is decoupled. We give mass to the flavor $N_f$ by adding a mass term:
\begin{eqnarray}
W'=W+yM^{N_f}_{N_f}.
\label{dec885774}
\end{eqnarray}
Then one can write directly the form of the matrices (without repeating known calculations known from the standard case):
\begin{eqnarray}
M=
\left(
\begin{array}{cc}
\tilde{M}&0\\
0&y.
\end{array}
\right).
\label{fomrm7}
\end{eqnarray}
Moreover:
\begin{eqnarray}
M'=
\left(
\begin{array}{cc}
(N_f-1)\tilde{M}^{\prime}y&0\\
0&y_2
\end{array}
\right).
\label{fomrmofmprime}
\end{eqnarray}
Here we needed to take into account a decoupling limit that preserves the general expression for the field $M'$. Then one can write the relations between the old and the new invariants for $N_f-1$ flavors:
\begin{eqnarray}
&&I_1=\tilde{I_1}y
\nonumber\\
&&I_2=(N_f-1)\tilde{I_2}y+yy_2
\nonumber\\
&&I_3=(N_f-1)^{N_f-1}y^{N_f-1}\tilde{I_3}y_2.
\label{expr657888}
\end{eqnarray}
Then,
\begin{eqnarray}
&&W=(N-N_f)(I_1t-I_2)^{-\frac{N_f}{N-N_f}}I_3^{\frac{1}{N-N_f}}(\Lambda^b)^{\frac{1}{N-N_f}}\times
\nonumber\\
&&(-1)^{\frac{N_f}{N-N_f}}N_f^{\frac{N_f}{N-N_f}}=
\nonumber\\
&&[\tilde{I}_1yt-(N_f-1)\tilde{I}_2y-yy_2]^{-\frac{N_f}{N-N_f}}\tilde{I}_3^{\frac{1}{N-N_f}}(N_f-1)^{\frac{N_f-1}{N-N_f}}y_2^{\frac{1}{N-N_f}}y^{\frac{N_f-1}{N-N_f}}\times
\nonumber\\
&&(-1)^{\frac{N_f}{N-N_f}}N_f^{\frac{N_f}{N-N_f}}=
\nonumber\\
&&X_2y^{\frac{1}{N_f-N}}.
\label{exprfinal64554}
\end{eqnarray}
The decoupling conditions are:
\begin{eqnarray}
&&\frac{\partial W}{\partial y_2}=\frac{\partial W'} {\partial y_2}=0
\nonumber\\
&&\frac{\partial W'}{\partial y}=\frac{\partial W}{\partial y}+m=0.
\label{twoexpr774665}
\end{eqnarray}
Then the  first equation in Eq. (\ref{twoexpr774665}) leads to:
\begin{eqnarray}
y_2=-\tilde{I_1}(N_f-2)!+\tilde{I_2}.
\label{firstrez664553}
\end{eqnarray}
The second equation in Eq. (\ref{twoexpr774665}) yields:
\begin{eqnarray}
&&W'=\frac{N+1-N_f}{N-N_f}W
\nonumber\\
&&y^{\frac{1}{N_f-N}}=X_2^{\frac{1}{N_f-1-N}}m^{-\frac{1}{N_f-1-N}}(N-N_f)^{-\frac{1}{N_f-N-1}}.
\label{rez663552}
\end{eqnarray}
By introducing $y$ and $y_2$ from Eqs. (\ref{firstrez664553}) and (\ref{rez663552}) into Eq. (\ref{exprfinal64554}) and after some calculations one obtains:
\begin{eqnarray}
W'=(N_f+1-N)(N_f-1)^{\frac{N_f-1}{N-N_f+1}}(-1)^{\frac{N_f-1}{N-N_f+1}}[\tilde{I}_1(N_f-2)!-\tilde{I_2}]^{-\frac{N_f-1}{N+1-N_f}}\tilde{I}_3^{\frac{1}{N+1-N_f}}.
\label{finalrez775664}
\end{eqnarray}

\section{Dynamical supersymmetry breaking and the new superpotential}

First we note that requiring that the massless degree of freedom satisfy some constraint is equivalent with the introduction of a large mass term for the same degrees of freedom. In our approach the main degrees of freedom were the field $M'$ such that  it makes sense to introduce a large mass term for ($M$) (Alternatively one may consider $M$ the main degrees of freedom with the same main results). The new superpotential will be:
\begin{eqnarray}
W_1=W +m^j_iM^i_j.
\label{newsup76885}
\end{eqnarray}
We are interested in studying the vacuum. For that one needs to solve:
\begin{eqnarray}
&&\frac{\partial W_1}{\partial M^{i}_j}=
H(-\frac{N_f}{N-N_f})[\frac{I_1t}{M^j_i}-M^{\prime j_i}][I_1t-I_2]^{-\frac{N_f}{N-N_f}-1}I_3^{\frac{1}{N-N_f}}+m^j_i=0
\nonumber\\
&&\frac{\partial W_1}{\partial M^{\prime i}_j}=H\frac{1}{N-N_f}[N_f\frac{M^j_i}{I_1t-I_2}+\frac{1}{M^{\prime j}_i}][I_1t-I_2]^{-\frac{N_f}{N-N_f}}I_3^{\frac{1}{N-N_f}}=0.
\label{twosolve6664553}
\end{eqnarray}

Now the set-up of the problem is as such the fields $M$ get eliminated. Note that $M$ and $M'$ are two arbitrary matrices that in general do not commute. This means that one can consider one of the matrices diagonal but cannot assume that the other is also diagonal. We will consider $M$ diagonal. Since all elements of $M$ have masses which in the end will be taken to infinity one can verify through direct calculations or by simple analogy with the behavior of the $W_{ADS}$ superpotential that the matrix $M$ must have all components equal to zero. Then the first equation in (\ref{twosolve6664553}) is clearly verified. One can write the second equation in (\ref{twosolve6664553}) as:
\begin{eqnarray}
[N_fM^j_i+(I_1t-I_2)\frac{1}{M^{\prime j}_i}][I_1t-I_2]^{-\frac{N_f}{N-N_f}-1}I_3^{\frac{1}{N-N_f}}=0.
\label{res64553}
\end{eqnarray}
The first element in the first  square bracket is zero. Then the second must  also be zero:
\begin{eqnarray}
\frac{1}{M^{\prime j}_i}[I_1t-I_2]^{-\frac{N_f}{N-N_f}}I_3^{\frac{1}{N-N_f}}=0.
\label{seceq4243}
\end{eqnarray}
The equation above must be true for both diagonal and off diagonal elements of $M'$. Then it is necessary that:
\begin{eqnarray}
\frac{1}{M^{\prime j}_i}I_3^{\frac{1}{N-N_f}}=0.
\label{result56647}
\end{eqnarray}
Consider $I_3=\infty$. We assume without loss of generality that in order for the determinant to be zero (and by the power of symmetry) either the diagonal elements are infinite and the off-diagonal elements are zero or the reciprocal. Then one must have for the first case $\frac{1}{N-N_f}-\frac{1}{F}<0$ and also $\frac{1}{N-N_f}+\frac{1}{N_f}<0$. The reversed case leads to the same conclusion. It is clear that these conditions cannot be satisfied so $I_3$ cannot be infinite. Then we assume $I_3=0$. The following conditions must be satisfied for this case: $\frac{1}{N-N_f}-\frac{1}{F}>0$ and $\frac{1}{N-N_f}+\frac{1}{N_f}>0$. These constraints are satisfied for
$N_f>\frac{N}{2}$. There is no other solution. Then one can tune the other eigenvalues accordingly to satisfy both equations in (\ref{twosolve6664553}).

We conclude that for $N_f>\frac{N}{2}$ there is a  zero of the potential for finite values of the fields $M'$ and thus supersymmetry is not dynamically broken. For $N_f\leq\frac{N}{2}$ there is no zero of the potential even for infinite vacuum expectation values so supersymmetry is clearly broken.

\section{Conclusions}

In \cite{ADS}, \cite{Seiberg1}, \cite{Seiberg2} it was shown that for supersymmetric QCD with $N_f<N$ flavors the low energy regime is described by the $ADS$ superpotential. Considering all the anomalies and charges this potential is unique. However there is a slew of superpotentials that one can construct if one introduces additional degrees of freedom. Of course the additional degrees of freedom come with constraint equations or with large mass terms.
Here we introduced one single additional field $M^{\prime i}_j$ described in Eq. (\ref{res534428}) and two possible invariants that contain this field. Then we were able to build a consistent superpotential out of these degrees of freedom.

One might wonder: why reconsider a matter already settled? The reason is that properties of the theory that are not manifest in terms of some set of variables may become evident in terms of a different one. In our case the clear determination of the situations when supersymmetry is dynamically broken and when it is not was at the core of all our endeavours.

In the present work we showed that for $N_f>\frac{N}{2}$ the theory displays a fully supersymmetric vacuum for finite values of the fields therefore there is no dynamical breaking of the supersymmetry.  For $N_f\leq\frac{N}{2}$ there is no value of the fields, finite or infinite for which the scalar potential is zero. Consequently supersymmetry is dynamically broken. Thus in terms of the new variables the unwanted  behavior of the $W_{ADS}$ superpotential where the vacuum is obtained for fields at infinity is not encountered.

One can introduce  new version of the superpotential in terms of additional degrees of freedom also for the cases with $N_f\geq N$. But this is a topic of discussion for another work.

\end{document}